\documentclass[aps,prl,reprint,showpacs,superscriptaddress]{revtex4-1}

\usepackage[dvips]{graphicx}
\begin{document}

\title{Nanoscale Dichotomy of Ti 3\textit{d} Carriers\\ Mediating the Ferromagnetism in Co:TiO$_2$ Anatase Thin Films}

\author{T.~Ohtsuki}
\affiliation{RIKEN SPring-8 Center, Sayo-cho, Sayo-gun, Hyogo 679-5198, Japan}
\author{A.~Chainani}
\affiliation{RIKEN SPring-8 Center, Sayo-cho, Sayo-gun, Hyogo 679-5198, Japan}
\author{R.~Eguchi}
\affiliation{RIKEN SPring-8 Center, Sayo-cho, Sayo-gun, Hyogo 679-5198, Japan}
\affiliation{Institute for Solid State Physics, The University of Tokyo, Kashiwa, Chiba 277-8581, Japan}
\author{M.~Matsunami}
\affiliation{RIKEN SPring-8 Center, Sayo-cho, Sayo-gun, Hyogo 679-5198, Japan}
\affiliation{Institute for Solid State Physics, The University of Tokyo, Kashiwa, Chiba 277-8581, Japan}
\author{Y.~Takata}
\affiliation{RIKEN SPring-8 Center, Sayo-cho, Sayo-gun, Hyogo 679-5198, Japan}
\author{M.~Taguchi}
\affiliation{RIKEN SPring-8 Center, Sayo-cho, Sayo-gun, Hyogo 679-5198, Japan}
\author{Y.~Nishino}
\affiliation{RIKEN SPring-8 Center, Sayo-cho, Sayo-gun, Hyogo 679-5198, Japan}
\author{K.~Tamasaku}
\affiliation{RIKEN SPring-8 Center, Sayo-cho, Sayo-gun, Hyogo 679-5198, Japan}
\author{M.~Yabashi}
\affiliation{JASRI/SPring-8, Sayo-cho, Sayo-gun, Hyogo 679-5198, Japan}
\author{T.~Ishikawa}
\affiliation{RIKEN SPring-8 Center, Sayo-cho, Sayo-gun, Hyogo 679-5198, Japan}
\author{M.~Oura}
\affiliation{RIKEN SPring-8 Center, Sayo-cho, Sayo-gun, Hyogo 679-5198, Japan}
\author{Y.~Senba}
\affiliation{JASRI/SPring-8, Sayo-cho, Sayo-gun, Hyogo 679-5198, Japan}
\author{H.~Ohashi}
\affiliation{JASRI/SPring-8, Sayo-cho, Sayo-gun, Hyogo 679-5198, Japan}
\author{S.~Shin}
\affiliation{RIKEN SPring-8 Center, Sayo-cho, Sayo-gun, Hyogo 679-5198, Japan}
\affiliation{Institute for Solid State Physics, The University of Tokyo, Kashiwa, Chiba 277-8581, Japan}

\date{\today}

\begin{abstract}
We study the surface and bulk electronic structure of the room-temperature ferromagnet Co:TiO$_2$ 
anatase films using soft and hard x-ray photoemission spectroscopy with probe sensitivities of 
$\sim$1 nm and $\sim$10 nm, respectively. We obtain direct evidence of metallic Ti$^{3+}$ states 
in the bulk, which get suppressed to give a surface semiconductor, thus indicating a surface-bulk 
dichotomy. X-ray absorption and high-sensitivity resonant photoemission spectroscopy reveal 
Ti$^{3+}$ electrons at the Fermi level (E$_F$) and high-spin Co$^{2+}$ electrons occurring away 
from E$_F$. The results show the importance of the charge neutrality condition: 
Co$^{2+}$ + V$_{O}$$^{2-}$ + 2Ti$^{4+}$ $\leftrightarrow$ Co$^{2+}$ + 2Ti$^{3+}$ (V$_O$ is oxygen 
vacancy), which gives rise to the elusive Ti 3\textit{d} carriers mediating ferromagnetism via 
the Co 3\textit{d}-O 2\textit{p}-Ti 3\textit{d} exchange interaction pathway of the occupied orbitals.
\end{abstract}


\maketitle

Dilute ferromagnetic oxides have been at the forefront of a paradigm shift in the search for 
spintronics and magneto-optic device materials~\cite{citenum01,citenum02,citenum03,citenum04,
citenum05,citenum06}. However, for achieving 
reliable thin-film device performance, it is necessary to ensure the surface and bulk electronic 
properties of candidate materials. This is particularly true for high Curie-temperature oxide 
ferromagnets for spintronics which can be synthesized only in thin film form, such as 
Co:TiO$_2$~\cite{citenum02}, Mn:ZnO~\cite{citenum03}, Cr:In$_2$O$_3$~\cite{citenum04}, etc. The 
main question that still remains enigmatic: What changes drive the long-range coupling of local 
moments in an insulating parent? The pre-runners, namely, dilute magnetic semiconductors (DMSs) 
which consisted of magnetic 3\textit{d} transition metal (TM) ions doped in III-V and II-VI 
compounds, showed a Curie-temperature (\textit{T}$_c$) well-below room temperature e.g.~InMnAs, 
GaMnAs, ZnCrTe, etc~\cite{citenum07,citenum08,citenum09}. Theoretical studies of carrier-mediated 
ferromagnetism played a pivotal role in predicting high-\textit{T}$_c$ ferromagnets in the dilute 
substitution limit~\cite{citenum10}. The spin-charge degrees of freedom and the anisotropic character 
of the dopant TM \textit{d}-orbitals results in a directional dependence of exchange energies and 
\textit{T}$_c$'s of upto 170\ K~\cite{citenum11,citenum12}. The limitation of \textit{T}$_c$'s 
significantly lower than room temperature motivated studies on alternative parents, like oxides and 
nitrides. The discovery of ferromagnetism in Co-doped TiO$_2$ (Co:TiO$_2$), with a \textit{T}$_c$ 
exceeding 300\ K~\cite{citenum02}, was crucial in expanding the field to oxides, leading to a rapid 
increase of new materials and phenomena arising from a synergetic marriage of semiconductor physics 
and strongly correlated systems.

From extensive magnetic and transport studies, it is now well-accepted that Co:TiO$_2$ films grown in 
a reducing environment or annealed in vacuum result in Co segregation~\cite{citenum13,citenum14}, 
while films grown in an oxidizing condition are intrinsically ferromagnetic with high spin 
Co$^{2+}$~\cite{citenum15,citenum16}. While initial x-ray magnetic circular dichroism (XMCD) 
measurements and annealing in vacuum concluded that the ferromagnetism was extrinsic due to segregated 
Co metal clusters~\cite{citenum14}, subsequent XMCD studies showed intrinsic ferromagnetism in 
Co:TiO$_2$ grown in oxidizing conditions~\cite{citenum17}. A soft x-ray surface-sensitive photoemission 
study indicated the importance of exchange interaction between the occupied Co \textit{t}$_{2g}$ and 
conduction band Ti \textit{t}$_{2g}$ states~\cite{citenum18}, but concluded that anatase Co:TiO$_2$ are 
wide band gap semiconductors. Recent experimental studies have emphasized the importance of oxygen 
vacancies (V$_O$) in Co:TiO$_2$~\cite{citenum15,citenum19}. The above results are consistently explained 
by the theoretical model of a hydrogenic spin-split donor-impurity band interacting via ferromagnetic 
exchange with the localized spins to realize high-\textit{T}$_c$ ferromagnetism~\cite{citenum20}. 
However, the experimental results are still not conclusive about the origin of the donor band: whether 
it is derived from shallow oxygen vacancies or cation vacancies, singly occupied vacancies or doubly 
occupied vacancies, or due to Co$^{2+}$-V$_O$ complexes~\cite{citenum19}. This is a critical missing 
input to the mechanism of ferromagnetism in Co:TiO$_2$. In this work, we address this issue using 
surface and bulk-sensitive~\cite{citenum21} photoemission spectroscopy (PES), x-ray absorption (XAS) and 
high-sensitivity resonant(R) PES. Our results unambiguously show that the carriers are the occupied 
Ti$^{3+}$ 3\textit{d} electrons, a possibility completely neglected to date, and these electrons exhibit 
a nanoscale surface-bulk dichotomy of the electronic structure of Co:TiO$_2$ which would play a crucial 
role in real device applications.

Co:TiO$_2$ thin films were fabricated on SrTiO$_3$(100) single crystal substrates by pulsed laser 
deposition (PLD) method. A KrF excimer laser (Lambda Physik, $\lambda$ = 248\ nm) was used to ablate 
sintered pellet target (Ti$_{0.95}$Co$_{0.05}$O$_2$). The laser fluence and repetition rate were set to 
1.2\ J/cm$^2$/pulse and 3 Hz, respectively. During deposition, the oxygen partial pressure 
(\textit{P}(O$_2$)) and substrate temperature were fixed at 1 $\times$ 10$^{-6}$ Torr and 
650\ ${}^\circ\mathrm{C}$, respectively. X-ray diffraction (XRD) measurements confirmed the epitaxial 
growth of these films in (001)-oriented anatase phase (See \cite{supplement} for details of 
characterization), which also showed ferromagnetism at room temperature~\cite{citenum06}.

Hard x-ray(HAX)-PES measurements were carried out at undulator beamline BL29XU at SPring-8 using a 
photon energy of 7940\ eV and Gammadata-Scienta R4000-10kV electron energy analyzer~\cite{citenum21}. 
The total energy resolution was set to 250\ meV. The soft x-ray(SX)-PES measurements were carried out 
at undulator beamline BL17SU of SPring-8. The PES spectra were obtained using a Gammadata-Scienta 
SES-2002 electron energy analyzer. The total energy resolutions were set to 240\ meV with $h\nu$ = 1200\ eV 
and 230\ meV for RPES measurements, respectively. The Fermi level (E$_F$) of the samples was referred to 
that of an Au film deposited on the sample substrate. The XAS spectra were taken in the total electron 
yield mode and were calibrated by measuring Au 4\textit{f} PES peak excited with first- and second-order 
light. All the measurements were performed at room temperature.

 \begin{figure}
 \includegraphics[width=8.6cm, clip]{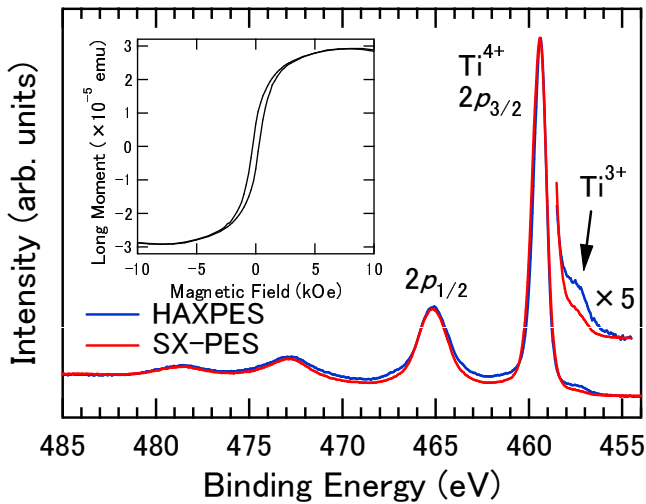}%
 \caption{\label{figure1}Ti 2\textit{p} core-level PES of Co:TiO$_2$ thin film by HAXPES (blue) and SX-PES 
(red), respectively. HAXPES and SX-PES were measured with photon energies of 7940\ eV and 1200\ eV, 
respectively. The close up view of the spectra ($\times$5) at lower binding energy side of main Ti 
2\textit{p}$_{3/2}$ peak is shown together. Ti$^{3+}$components can be clearly recognized. The inset is 
the magnetization vs.~magnetic field curve of the same film measured at room temperature. The observed 
hysteresis confirms the room temperature ferromagnetism in this Co:TiO$_2$ thin film.}
 \end{figure}

Figure \ref{figure1} shows the Ti 2\textit{p} core-level spectra of Co(5\ at\%):TiO$_2$ thin films grown 
epitaxially on a SrTiO$_3$(100) substrates. The inset shows the hysteresis at 300\ K confirming the 
ferromagnetism~\cite{supplement}. The Ti 2\textit{p} core-level spectra measured with SX-PES 
($h\nu$ = 1200\ eV) and HAXPES ($h\nu$ = 7940\ eV) show a main Ti 2\textit{p}$_{3/2}$ peak at 459.4\ eV 
and 2\textit{p}$_{1/2}$ peak at 465.2\ eV due to 
Ti$^{4+}$ states, accompanied by charge-transfer satellites at 13\ eV higher binding 
energies~\cite{citenum22}. In addition, a fine structure is observed at 1.8\ eV lower binding energy to 
the main peak. We plot the SX-PES and HAXPES on an enlarged ($\times$5) scale and normalized to the main 
peak intensity (inset Fig. \ref{figure1}), which clearly shows that this feature is enhanced in the 
HAXPES data compared to SX-PES. This feature is not observed in the case of stoichiometric Ti$^{4+}$ 
(\textit{d}$^0$ state) titanates, but shows up in reduced/electron-doped titanates indicative of Ti$^{3+}$ 
states~\cite{citenum23,citenum24}. A similar behaviour was recently found for the 2D-electron gas at 
SrTiO$_3$-LaAlO$_3$ interface using HAXPES~\cite{citenum25}. The present results indicate that Ti$^{3+}$ 
components in Co:TiO$_2$ thin films are clearly present in bulk-sensitive HAXPES with a probing depth 
of $\sim$10\ nm and they are suppressed in the top 1-2\ nm, as probed by surface sensitive SX-PES. 
For $\delta$ oxygen vacancies, 2$\delta$ electrons are introduced into valence band states consisting of Ti 
3\textit{d} and O 2\textit{p} electrons~\cite{citenum26}. In addition, if Co is doped as Co$^{2+}$ (as 
confirmed below), oxygen defects are also necessary to keep charge neutrality. If the carriers are trapped, 
these are described as Co$^{2+}$ + V$_O$$^{2-}$ complexes~\cite{citenum19}. However, our core-level O 
1\textit{s} spectra show a clean single peak \cite{supplement} but with a small shift in binding 
energies, suggestive of negligible role of oxygen vacancies. Taken together with the presence of Ti$^{3+}$, 
it indicates the following charge neutral transformation, Co$^{2+}$ + V$_{O}$$^{2-}$ + 2Ti$^{4+}$ 
$\leftrightarrow$ Co$^{2+}$ + 2Ti$^{3+}$. This is consistent with no changes in Co co-ordination using 
extended x-ray absorption fine structure (EXAFS), suggesting that the defects are mobile and not localized 
to the first co-ordination shell of Co$^{2+}$ ions~\cite{citenum15}.

\begin{figure}
 \includegraphics[width=8.6cm, clip]{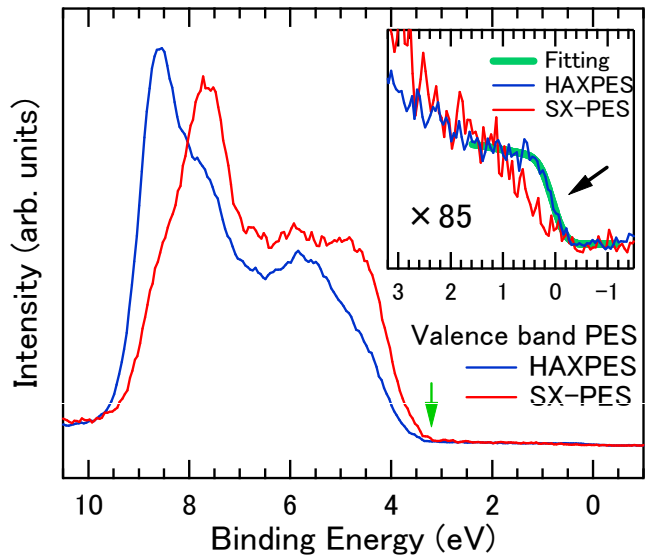}%
 \caption{\label{figure2}Valence band PES of Co:TiO$_2$ thin film by HAXPES (blue) and SX-PES (red). The 
spectra are normalized by the peak area. A green arrow indicates the position of reported band-gap 
value~\cite{citenum27}. The inset is the enlargement of the spectra ($\times$85) near-E$_F$. A green line 
is the fitting curve of HAXPES spectrum. HAXPES spectrum clearly shows Fermi-edge step (marked by black 
arrow).}
 \end{figure}

In order to confirm Ti$^{3+}$, we measured the valence band spectra with SX-PES and HAXPES 
(Fig. \ref{figure2}). Non-doped anatase TiO$_2$ is a wide-gap semiconductor with a band gap of 
3.2\ eV~\cite{citenum27}, and the observed gap in Fig. \ref{figure2} seems to confirm it (green arrow in 
Fig.~\ref{figure2}). This indicates that E$_F$ is pinned at the bottom of the conduction band. However, 
in the inset of Fig. \ref{figure2}, we plot high signal-to-noise ratio near-E$_F$ data on an enlarged 
($\times$85) y-scale, and it clearly shows in-gap states. More importantly, the spectra show a depletion 
of states at E$_F$ for the SX-PES data while the HAXPES data shows a clear Fermi-edge step. This 
indicates a nanoscale surface-semiconductor : bulk-metal dichotomy. Next, we investigate the electronic 
character of the in-gap region in detail, using Co 2\textit{p}-3\textit{d} and Ti 2\textit{p}-3\textit{d} 
XAS and RPES measurements which provides unoccupied and occupied element/site-projected and 
orbital-selective 3\textit{d} partial density of states (DOS), respectively.

\begin{figure}
 \includegraphics[width=8.6cm, clip]{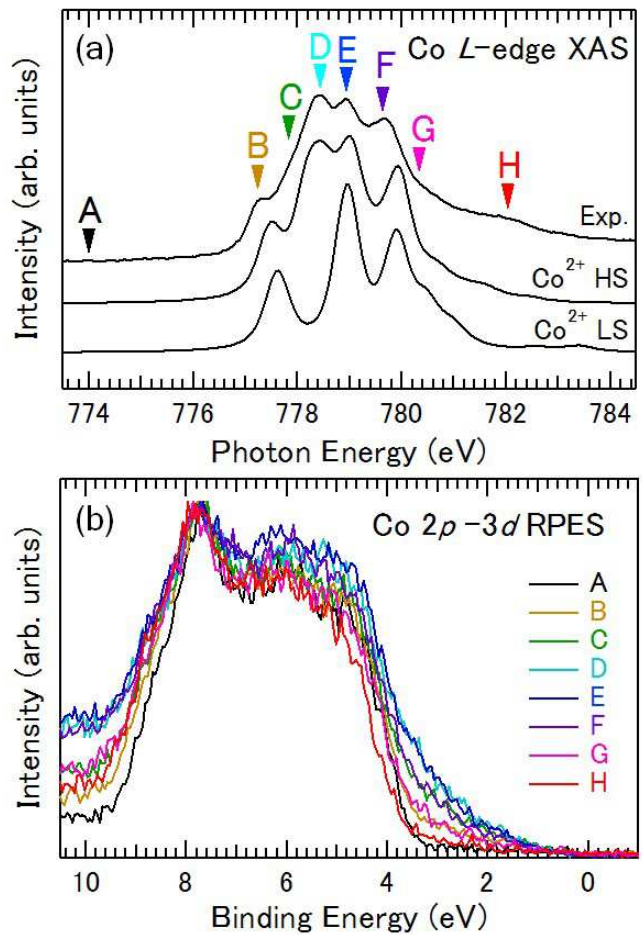}%
 \caption{\label{figure3}(a) The XAS spectra of Co:TiO$_2$ thin film near Co \textit{L}$_3$-edge. The 
labels Exp., Co$^{2+}$ HS, and Co$^{2+}$ LS correspond to the experimental data, calculation of divalent 
high-spin state, and divalent low-spin state, respectively. (b) RPES spectra by Co 2\textit{p}-3\textit{d} 
resonance. The intensities of RPES spectra are normalized by peak height. The labels A-H correspond to 
excitation photon energies as marked in (a).}
 \end{figure}

Figure \ref{figure3}(a) shows the experimental XAS spectrum of anatase Co:TiO$_2$ at the 
Co \textit{L}$_3$-edge, compared with the theoretically calculated low-spin and high-spin Co$^{2+}$ spectrum 
of Co ions \cite{supplement}. The rich structure found in the experimental spectrum of 
Fig.~\ref{figure3}(a) shows very good correspondence with the calculated high-spin Co$^{2+}$ spectrum and 
also rules out the low spin Co$^{2+}$ configuration. It is also quite similar to that of rutile 
Co:TiO$_2$~\cite{citenum17}. The results thus confirm a high-spin Co$^{2+}$ ion substituted in the 
Ti$^{4+}$ site for Co:TiO$_2$.

Figure \ref{figure3}(b) shows the Co 2\textit{p}-3\textit{d} valence band RPES of Co:TiO$_2$ obtained with 
photon energies A-H marked in Fig.~\ref{figure3}(a). A comparison of the off-resonance spectrum (A) with 
on-resonance spectra (B-H) indicates that the spectral changes observed at $\sim$1 
to $\sim$3.5\ eV binding energy corresponds to the new Co 3\textit{d} states obtained in the 
gap, while changes in the O 2\textit{p} states result from hybridization changes accompanying the doping. 
From Fig.~\ref{figure3}(b), we also verify that the Co 3\textit{d} states are not located at E$_F$, thus 
indicative of localized Co 3\textit{d} electrons.

\begin{figure}
 \includegraphics[width=8.6cm, clip]{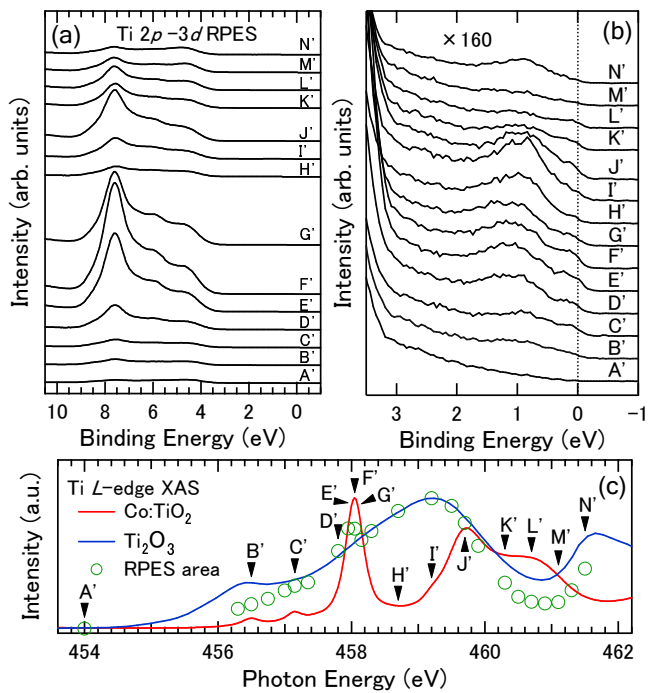}%
 \caption{\label{figure4}(a) Ti 2\textit{p}-3\textit{d} RPES spectra of Co:TiO$_2$ thin film plotted in wide 
region. The photoemission intensities are normalized by photon flux. (b) The enlarged scale ($\times$160) 
plot of (a) near-E$_F$. The labels A'-H' correspond to excitation photon energies as marked in (c). (c) The 
XAS spectra of Co:TiO$_2$ thin flim (red) and Ti$_2$O$_3$ single crystal~\cite{citenum29} (blue) around 
Ti \textit{L}$_3$-edge. The integrated peak areas from -0.5 to 2.5\ eV in (b) are also plotted by green 
open circle.}
 \end{figure}

Figure \ref{figure4} shows the Ti 2\textit{p}-3\textit{d} XAS and RPES of Co:TiO$_2$ thin films measured to 
clarify the Ti 3\textit{d} states in the electronic structure. The wide energy scale RPES spectra are shown 
in Fig.~\ref{figure4}(a) while the near-E$_F$ in-gap region spectra are shown in Fig.~\ref{figure4}(b). The 
photoemission intensities are normalized by photon flux and the high-sensitivity near-E$_F$ spectra are 
magnified ($\times$160) in order to characterize the spectral features. Off-resonant spectrum 
($h\nu$ = 454\ eV) shows no striking changes from the valence band PES shown in Fig.~\ref{figure2}. In 
contrast, the on-resonant spectra show two clear features: a weak feature at E$_F$ and another higher 
intensity feature at $\sim$1\ eV binding energy, arising from coherent and incoherent features 
of \textit{d}$^1$ Ti$^{3+}$ states in the Mott-Hubbard picture of strongly correlated 
electrons~\cite{citenum28}. Interestingly, we have plotted the integrated intensity of the in-gap states as 
a function of incident photon energy in Fig.~\ref{figure4}(c), and at first glance, it shows no 
correspondence with sharp features in the Ti 2\textit{p}-3\textit{d} XAS spectrum. However, a comparison 
with the Ti 2\textit{p}-3\textit{d} XAS of Ti$_2$O$_3$ (Ti$^{3+}$) single crystal fractured in vacuum 
(blue line in Fig.~\ref{figure4}(c)~\cite{citenum29}) shows a very good match with the peak at 459.2\ eV. 
Hence, the resonant enhancement around 459\ eV is due to the Ti$^{3+}$ component in Co:TiO$_2$.

Thus, while the core level and valence band spectra of Figs.~\ref{figure1} and \ref{figure2} clearly show 
Ti$^{3+}$ states, the RPES results show that the in-gap states (inset of Fig.~\ref{figure2}) are actually 
derived from Co 3\textit{d} and Ti 3\textit{d} states. The magnetic Co 3\textit{d} states are spread over 
a wide energy range from $\sim$1 to 3.5\ eV, while Ti 3\textit{d} states which provide carriers 
occur from E$_F$ to about 2\ eV. Consequently, the partial DOS of Co 3\textit{d} and Ti 3\textit{d} overlap 
significantly and would be well-hybridized, constituting a Co 3\textit{d}-O 2\textit{p}-Ti 3\textit{d} 
ferromagnetic exchange pathway responsible for a spin-charge coupled nanoscale dichotomy. The results 
conclusively indicate that the ferromagnetism of Co:TiO$_2$ is explained by a carrier-induced mechanism 
involving Ti 3\textit{d} states, consistent with a donor exchange mechanism proposed earlier~\cite{citenum19}, 
but with the occupied Ti 3\textit{d} orbitals playing the role of the donors.

\begin{acknowledgments}
The measurements of HAXPES and SX-PES were carried out with the approval of the 
RIKEN SPring-8 Center (Proposal No.~20090041 and No.~20090033, respectively).
\end{acknowledgments}

\bibliography{bibfile}

\end{document}